\pdfoutput=1
\documentclass[aps,prb,twocolumn,showpacs,longbibliography,superscriptaddress]{revtex4-1}
\usepackage{graphicx}
\usepackage{bm}
\usepackage{amsmath}
\usepackage{amssymb}
\usepackage{euscript}
\usepackage{verbatim}
\usepackage{setspace}
\usepackage{xcolor}
\usepackage{amsfonts}
\usepackage{braket}
\usepackage{subfigure}

\newcommand{\be}{\begin{equation}}
\newcommand{\ee}{\end{equation}}
\newcommand{\bea}{\begin{eqnarray}}
\newcommand{\eea}{\end{eqnarray}}

\newcommand{\Renyi}[0]{R\'{e}nyi~}
\allowdisplaybreaks

\newcommand{\Tr}[1]{\mathrm{Tr} #1}

\begin{document}

\title{Symmetric inseparability and number entanglement in charge conserving mixed states}
\author{Zhanyu Ma}
\affiliation{School of Physics and Astronomy, Tel Aviv University, Tel Aviv 6997801, Israel}

\author{Cheolhee Han}
\affiliation{School of Physics and Astronomy, Tel Aviv University, Tel Aviv 6997801, Israel}

\author{Yigal Meir}
\affiliation{Department of Physics, Ben-Gurion University of the Negev, Beer Sheva 84105, Israel}
\affiliation{The Ilse Katz Institute for Nanoscale Science and Technology, Ben-Gurion University of the Negev, Beer Sheva 84105, Israel}

\author{Eran Sela}
\affiliation{School of Physics and Astronomy, Tel Aviv University, Tel Aviv 6997801, Israel}

\begin{abstract}
	We explore sufficient conditions for inseparability in mixed  states with  a globally conserved charge, such as a particle number. We argue that even separable states may contain entanglement in fixed charge sectors, as long as the state can not be separated into charge conserving components. As a witness of symmetric inseparability 
	we study the number entanglement (NE), $\Delta S_m$, defined as the entropy change due to a subsystem's charge measurement. Whenever $\Delta S_m>0$, there exist inseparable charge sectors, having finite  (logarithmic) negativity, even when the full state is either separable or has vanishing negativity. 
	We demonstrate that the NE is not only a witness of symmetric inseparability, but also an entanglement monotone. Finally, we study the scaling of $\Delta S_m$ in thermal 1D  systems combining high temperature expansion and conformal field theory. 
\end{abstract}

\maketitle

\section{Introduction} 
Quantifying and exploiting quantum entanglement is a central activity unifying quantum information and condensed matter~\cite{calabrese2004entanglement,amico2008entanglement,horodecki2009quantum,eisert2010colloquium,laflorencie2016quantum}. Entanglement quantifies the inseparability of quantum states. A  separable state can be written as
\be
\label{eq:separability}
\rho=\sum_i p_i \rho_A^i \otimes  \rho_B^i,
\ee
where $\rho_A^i$ and $\rho_B^i$ are density matrices for the two subsystems $A$ and $B=\bar{A}$, with probabilities $\sum_i p_i=1$. Determining whether a mixed state is separable is NP hard~\cite{gurvits2003classical,gharibian2008strong,qian2020separability}. 
The criterion for a state to have finite negativity, i.e. negative eigenvalues after  partial transposition, 
is a sufficient condition for inseparability~\cite{Peres1996,vidal2002computable,plenio2005logarithmic}. 

In this work we consider mixed states in the presence of a conserved charge, such as the total particle number  $\hat{N}=\hat{N}_A+\hat{N}_B$, such that $[\rho,\hat{N}]=0$.  The separability condition Eq.~(\ref{eq:separability}) does not require that the individual classically combined components $\rho_A^i \otimes  \rho_B^i$ are symmetric. Since the latter should describe plausible physical states, we supplement Eq.~(\ref{eq:separability}) by the symmetry condition
\be
\label{symmetrycondition}
\forall_i~~~ [ \rho_A^i \otimes  \rho_B^i,\hat{N}]=0.
\ee
As a familiar example, a state  of free bosons hoping on a lattice at temperature $T$ and chemical potential $\mu$, $\rho = e^{-\beta (H-\mu N)}/Z$, where $\beta = T^{-1}$ is the inverse temperature and $Z$ is the partition function
, has a separable form in terms of coherent states~\cite{lu2020structure}. But since coherent states are not number eigenstates, Eq.~(\ref{symmetrycondition}) is not satisfied. Thus while this state has zero negativity, it does not satisfy   symmetric-separability. As a consequences of symmetric-inseparability, we will show that such a state must contain entanglement in specific charge sectors.

\begin{figure}[b]
	\includegraphics[width=\linewidth]{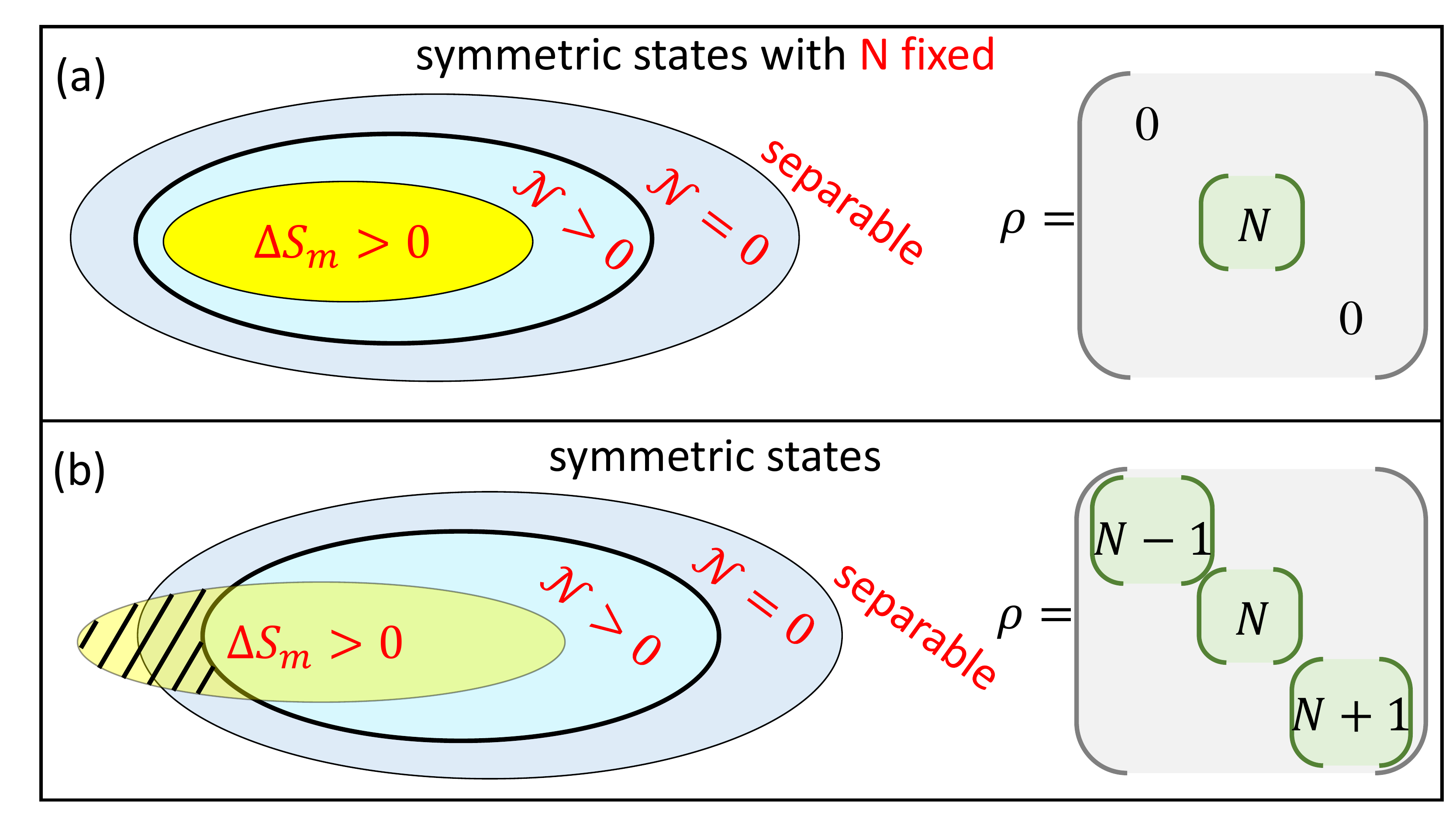}
	\caption{The space of symmetric states $\rho$ is divided by the thick ellipse into states with either finite- or zero logarithmic negativity $\mathcal{N}$. Part of the latter are separable (white region). (a) For fixed particle number $N$, the set of states with finite $\Delta S_m$ (yellow region) is included in the set of states with finite negativity. (b) For general states with multiple total-charge blocks, describing e.g. systems with a chemical potential, the set of states with finite $\Delta S_m$  may include separable states.
	}
	\label{fig:0}
\end{figure} 

In this paper we explore a  measurable quantity that provides a necessary condition for symmetric-separability, Eqs.~(\ref{eq:separability}) and (\ref{symmetrycondition}). We consider the entropy change due to  an unselective  measurement of the subsystem's charge,
\be
\label{eq:maindef}
\Delta S_m = S(\rho_m) - S(\rho),
\ee
where $\rho_m = \sum_{N_A} \Pi(N_A) \rho \Pi(N_A)$, $S(\rho) = -\rm{Tr} (\rho \log \rho)$,  and $\Pi(N_A)$ is a projector to a fixed number $N_A$ of the subsystem. Equivalently, $\Delta S_m = S(\rho|| \rho_m)$ is the relative entropy of charge coherence with respect to $N_A$~\cite{macieszczak2019coherence}; we refer to $\Delta S_m$ as the number entanglement (NE).

We show that $\Delta S_m$ vanishes for symmetrically separable states, so that $\Delta S_m > 0$ implies that a state can not be symmetrically separable. In systems with a fixed particle number, $\Delta S_m > 0$ may only occur when negativity is already present, see Fig.~\ref{fig:0}(a).  As our main result, in the general case with fluctuating number of particles, we show that whenever $\Delta S_m>0$, there exists some total charge-$N$ block in the density matrix which is inseparable and has finite negativity, see Fig.~\ref{fig:0}(b). Then entanglement can be extracted via a projection to a total charge sector.  Importantly, $\Delta S_m>0$ may happen even  in (non-symmetrically) separable states.  In these cases the only way to display separability is by violating the symmetry Eq.~(\ref{symmetrycondition}). 

Similar to Bell's inequalities, an entanglement witness gives a yes/no answer to the separability condition. Remarkably, we find that $\Delta S_m$ is not only an entanglement witness in mixed states with a conserved particle number. It is also an entanglement monotone. An entanglement monotone is non-increasing under local operations and classical communication (LOCC)~\cite{plenio2014introduction}. Here, we show that the NE, $\Delta S_m$, is non-increasing under symmetric LOCC,  which conserve the total charge. This gives a comparative meaning to the value of the NE, allowing an interpretation in terms of the number of Bell pairs in the charge sector.

The paper is organized as follows. In Sec.~\ref{se:properties} we describe the key properties of $\Delta S_m$ in general mixed states with a conserved charge. In Sec.~\ref{se:monotonicity} we show that $\Delta S_m$ is an entanglement monotone. In Sec.~\ref{se:negativity} 
we connect $\Delta S_m$ with negativity, showing that whenever $\Delta S_m>0$ there  exist a charge sector with finite negativity; examples are given in Sec.~\ref{se:examples}.  In Sec.~\ref{se:methods} we describe methods to compute $\Delta S_m$ in extended 1D systems, ranging from numerical free fermion methods, high temperature expansion, and conformal field theory.   We summarize in Sec.~\ref{se:summary}.


\section{General properties}
\label{se:properties} Consider general symmetric mixes states $\rho$ with $[\rho, \hat{N}]=0$. To demonstrate that $\Delta S_m$ is a measure of symmetric inseparability, we note that $\Delta S_m$ satisfies a number of properties:
\bea
&1.&~~~\Delta S_m \geq 0,\nonumber \\
&2.&~~~ \Delta S_m=0 {\rm{~for~ symmetric~separable~states}}.\nonumber
\eea
One can show that $\Delta S_m \ge 0$ by demonstrating that $\Delta S_m$  equals the relative entropy between $\rho$ and $\rho_m$~\cite{nielsen2002quantum,macieszczak2019coherence}
$S(\rho||\rho_{m})=\Tr{(\rho\log{\rho})}-\Tr{(\rho\log{\rho_{m}})}
$. This follows from the fact that
\begin{align*}
\Tr({\rho_{m}\log{\rho_{m}}})
&=\sum_{N_A}\Tr{\left (\Pi(N_A)\rho\Pi(N_A)\log{\rho_{m}}\right)}\\
&=\sum_{N_A}\Tr{\left (\rho \Pi(N_A)\log{(\rho_{m})}\Pi(N_A)\right)}\\
&=\Tr({\rho\log{\rho_m}}),
\end{align*}
where we used the property that $\rho_{m}$ commutes with all projection operators $\Pi(N_A)$. So $\Delta S_m \ge 0$ follows directly from the positivity of the relative entropy.

To show that $\Delta S_m=0$ for symmetrically separable states, satisfying Eqs.~(\ref{eq:separability}) and (\ref{symmetrycondition}), let us temporarily focus on the simple product state, $\rho=\rho_A\otimes \rho_B$. The requirement $[\rho, \hat{N}]=0$ implies $[\rho_A, \hat{N}_A]=0$ (by taking a partial trace on both sides of the equation). So $\rho_A$ actually commutes with all projection operators $\Pi(N_A)$. Thus the post-measurement state is the same as $\rho$ and there is no entropy change. The same is true for general symmetric-separable states since each component satisfies Eq.~(\ref{symmetrycondition}).

Additional useful properties we now show are: (3) $\Delta S_m$ is invariant under symmetry preserving local unitary transformations; (4) $\Delta S_m$ is invariant under the exchange of the roles of $A$ and $B$, and (5) $\Delta S_m$ is additive. 

3. \textit{Invariance under symmetry preserving local unitary transformations}: Here we consider local unitaries which preserve the total charge $N$. Because it is a local operator either acting in $A$ or in $B$, it preserves $N_A$, i.e. $[U,\hat{N}_A]=0$. These could be unitaries acting on internal degrees of freedom. Then it is not hard to show that
\bea
\Delta S_m & \to & S \left( \sum_{N_A} \Pi(N_A) U \rho U^\dagger \Pi(N_A)  \right) - S(U \rho U^\dagger) \nonumber \\
&=& S\left( \sum_{N_A} \Pi(N_A)  \rho  \Pi(N_A)  \right) - S( \rho ) = \Delta S_m,
\eea
where we used the properties $S(\rho) = S( U \rho U^\dagger)$ and $[\Pi(N_A),U]=0$ which follows from $[U,\hat{N}_A]=0$.

4. \textit{$\Delta S_m$ is symmetric if we exchange $A$ and $B$:} In other words if we choose to measure the particle number in $B$ or in $A$, the post measurement states are the same, as obtained by annihilating all off diagonal blocks with respect to $N_A$.

5. \textit{$\Delta S_m$ is additive}. Consider 2 flavors of particles  $\rho = \rho^{f_1} \otimes \rho^{f_2}$. Here we require that the number of particles of each flavor $N_{f_1}$ and $N_{f_2}$ are separately conserved. Now if we separately measure the particle number of both flavors of particles in subsystem $A$, it is straightforward to show that $\rho_m = \rho_{f_1,m} \otimes  \rho_{f_2,m}$ and hence $\Delta S_m(\rho)=S_m(\rho^{f_1})+S_m(\rho^{f_2})$.

\subsection{Pure states} We now study $S_m$ in pure states. The most general pure state with $N$ particles can be written as
\be
|\Psi \rangle = \sum_{N_A} \sqrt{P(N_A)} \left(\sum_{i,\alpha }c^{(N_A)}_{i,\alpha} |N_A,i \rangle_A  |N-N_A,\alpha \rangle_B  \right),
\ee
where $P(N_A)$ is the probability to find the subsystem $A$ with charge $N_A$, $i$ and $\alpha$ denote basis states in regions $A$ and $B$ for a given number of particles, and $c^{(N_A)}_{i,\alpha}$ are normalized as $\sum_{i,\alpha }|c^{(N_A)}_{i,\alpha}|^2=1$. For pure states obviously $S(\rho)=0$, and
\be
\Delta S_m = S(\rho_m)=-\sum_{N_A} P(N_A) \log P(N_A),~~{\rm{(pure~states)}}.
\ee 
Thus, $\Delta S_m$ coincides for pure states with the number entropy~\cite{goldstein2018symmetry,xavier2018equipartition,lukin2019probing,bonsignori2019symmetry,rakovszky2019entanglement,kiefer2020evidence,calabrese2021symmetry}, i.e. the entropy of the distrubution function of subsystem's charge. If one of the subsystems contains only one site (with no additional internal degrees of freedom) then the number entropy equals the entanglement entropy $S_{EE}=S(\rho_A)$ where $\rho_A = {\rm{Tr}}_B \rho$, but in general $S_{EE}>-\sum_{N_A} P(N_A) \log P(N_A)$. In addition we note that the number entropy is bounded from above by $\log(1+ N_{A,max}-N_{A,min})$ where $1+ N_{A,max}-N_{A,min}$ is the number of subsystem charge states. The part of the entanglement entropy not included in the number entropy is often referred to as configuration- or accessible entropy~\cite{wiseman2003entanglement,barghathi2018renyi,barghathi2019operationally}, and it admits a symmetry-resolution~\cite{goldstein2018symmetry, xavier2018equipartition, cornfeld2018imbalance, bonsignori2019symmetry, feldman2019dynamics, horvath2020symmetry, fraenkel2020symmetry, neven2021symmetry, fraenkel2021entanglement,murciano2021symmetry,vitale2021symmetry, azses2020identification,azses2020symmetry,azses2021observing},  see  Appendix~\ref{app:sre}. 

The relationship between $\Delta S_m$, entanglement, and number entropy, can be visualized by simple examples. Consider the state 
$\ket{\Psi}=\alpha\ket{01}+\beta\ket{10}$, 
defined on two sites, where $\ket{0}$ represents an empty site, $\ket{1}$ a filled site, with $|\alpha|^2+|\beta|^2=1$.
The reduced density matrix of the first site is $\rho_A=|\alpha|^2\ket{0}\bra{0}+|\beta|^2\ket{1}\bra{1}$. Consider measuring the particle number in the first site, yielding
\be
\rho_{m}=|\alpha|^2\ket{01}\bra{01}+|\beta|^2\ket{10}\bra{10}.
\ee
In this example where the subsystem consists of a single site, $N_A$ fully specifies the quantum state in $A$, and hence the entropy change coincides  with the entanglement entropy and number entropy.

Now consider the following state on 4  sites,
\be
\label{eq:phi}
\ket{\Phi}=\frac{1}{\sqrt{2}}(\ket{0101}+\ket{1010}),
\ee
which is entangled, and satisfies $[\hat{N},\ket{\Phi}\bra{\Phi}]=0$. 
Consider measuring the particle number in the first two sites. In this case the quantum state does not change, i.e., $\Delta S_m=0$. Similarly the number entropy vanishes. This example illustrates 
that $\Delta S_m$ 
does not capture the full entanglement, only the entanglement between different symmetry sectors.

However in general mixed states the number entropy is unrelated to entanglement. This can be seen by the following example of a symmetrically-separable (and hence unentangled) state,
\be
\rho = \sum_{N_A }P(N_A)  | N_A \rangle_A \langle N_A | \otimes   | N-N_A \rangle_B \langle N-N_A |.
\ee

Now consider instead the following state on 2 sites,
\be
\ket{\phi}=\ket{0}\otimes\frac{\ket{0}+\ket{1}}{\sqrt{2}},
\ee
which is clearly a product state, with no entanglement. However if we measure the particle number in the second site, the measured state is
\be
\rho_m=\ket{0}\bra{0}\otimes\frac{1}{2}(\ket{0}\bra{0}+\ket{1}\bra{1}),
\ee
and the entropy change is not zero. $\Delta S_m$ fails to indicate entanglement here because $[\hat{N},\ket{\phi}\bra{\phi}]\neq 0$. This example illustrates that we have to restrict to those states which possess a conserved quantity.

\section{Monotonicity} 
\label{se:monotonicity}
The dicussion so far emphasized, through properties 1-5 listed in Sec.~\ref{se:properties}, that the NE $\Delta S_m>0$ is a witness of symmetric inseparability in mixed states with a conserved charge. Yet, the actual value of $\Delta S_m$ did not play any role. Now we provide a comparative meaning to the value of $\Delta S_m$ in different states, showing that $\Delta S_m$ is actually an entanglement monotone in the presence of charge conservation. We apply the results of Ref.~\onlinecite{macieszczak2019coherence} which made related claims. 

To show that the NE is an entanglement monotone, we consider symmetric local operation and classical communication (LOCC) transformations,
\be
\mathcal{K}(\rho)=\sum_n K_n \rho K_n^\dagger,
\ee
where the Lindblad operators $K_n$ (i) satisfy~\cite{plenio2014introduction} $\sum_n K_n^\dagger K_n=I$, (ii) can be written as $K_n=K_n^{(A)} \otimes K_n^{(B)}$, and (iii) satisfy the symmetry condition $[K_n,\hat{N}]=0$.  This goes beyond unitary transformations, as it describes coupling to a bath, and also includes classical communication.   We will show that:
\bea
\Delta S_m{\rm{~does~not~increase~under~symmetric~LOCC}}.\nonumber
\eea
Thus the only way to increase $\Delta S_m$ is by genuine quantum entangling non-local operations. We demand the symmetry which is required for $\Delta S_m$ to be served as a witness of inseparability.

{\emph{Proof}}: The NE is defined as the relative entropy of the unmeasured and measured states, 
\be
\Delta S_m(\rho)= {\rm{Tr}}[\rho \log \rho]-{\rm{Tr}}[\rho \log \sum_{N_A} \Pi({N_A})\rho \Pi({N_A})   ].
\ee
We then need to prove that
\be
\label{eq:monotonicity}
\Delta S_m(\rho) \ge \Delta S_m(\mathcal{K}(\rho)),
\ee
where
\bea
\Delta S_m(\mathcal{K}(\rho))&=& {\rm{Tr}}[\mathcal{K}(\rho) \log \mathcal{K}(\rho)]\nonumber \\
&-&{\rm{Tr}}[\mathcal{K}(\rho) \log \sum_{N_A} \Pi({N_A})\mathcal{K}(\rho) \Pi({N_A})].
\eea
We proceed by showing that the symmetric LOCC transformation  commutes with the charge measurement, namely
\be
\label{eq:comuteSLOCC}
\sum_{N_A} \Pi({N_A})\mathcal{K}(\rho) \Pi({N_A})=\mathcal{K}\Big( \sum_{N_A} \Pi({N_A})\rho \Pi({N_A})\Big).
\ee
To prove this, we first use an alternative representation of the post-projective measurement state,
\be
\label{eq:14}
\rho_m=\sum_{N_A}\Pi({N_A})\rho\Pi({N_A})=\int_{-\pi}^{\pi}\frac{d\alpha}{2\pi}e^{i\alpha \hat{N}_A}\rho e^{-i\alpha \hat{N}_A}.
\ee
Notice that $\int_{-\pi}^{\pi}\frac{d\alpha}{2\pi}e^{i\alpha \hat{N}_A}\cdots e^{-i\alpha \hat{N}_A}$ acts as a projection operator, because $\int_{-\pi}^{\pi}\frac{dq}{2\pi}e^{iqN}=\delta_{N,0}$. In other words, it kills coherence between states with different $N_A$.
We also use the properties of the symmetric LOCC operators $K_n$. As shown in Ref.~\onlinecite{macieszczak2019coherence}, the $K_n$ operators satisfy, 
\be
\label{Ldelta}
[K_n,\hat{N}_A]=\delta_n K_n.
\ee 
We then say that the Lindblad operator $K_n$ has subsystem charge $\delta _n$.

Using Eqs.~(\ref{eq:14}) and (\ref{Ldelta}), we have
\begin{align}
&\sum_{N_A} \Pi({N_A})\mathcal{K}(\rho) \Pi({N_A})=\sum_n\int_{-\pi}^{\pi}\frac{d\alpha}{2\pi} e^{i\alpha\hat{N}_A}K_n \rho K_n^\dagger e^{-i\alpha\hat{N}_A}\nonumber\\
=&\sum_n\int_{-\pi}^{\pi}\frac{d\alpha}{2\pi} K_ne^{i\alpha\hat{N}_A} \rho  e^{-i\alpha\hat{N}_A}K_n^\dagger=\mathcal{K}\Big(\rho_m \Big).
\end{align}

As a result of Eq.~(\ref{eq:comuteSLOCC}), we see that the entropy change after the symmetric LOCC becomes $\Delta S_m(\mathcal{K}(\rho))=S(\mathcal{K}(\rho)||\mathcal{K}(\rho_m))$. Finally, we use a property of the relative entropy between two arbitrary density matrices $S(\rho||\sigma)$, being non-increasing under any completely positive trace preserving map applied on both $\rho$ and $\sigma$~\cite{nielsen2002quantum},
\be
\label{eq:newresult}
S(\rho||\sigma) \ge S(\mathcal{K}(\rho)||\mathcal{K}(\sigma)).
\ee
This proves the monotonicity condition Eq.~(\ref{eq:monotonicity})~\cite{plenio2014introduction}. 

We now discuss simple examples on two sites. First consider an initial product state $|00\rangle$ which is transformed via a local operation on the second site to the state $|\phi \rangle$ in Eq.~(\ref{eq:phi}). Under this LOCC  $\Delta S_m$  increases. But as explained above $|\phi \rangle$  does not commute with the symmetry, and then $\Delta S_m$ does not measure entanglement. This example emphasizes that $\Delta S_m$ can not increase under the specific LOCC transformations that  conserve charge, and explains why this non-increasing condition is restricted to symmetric LOCC.

As a second example, we consider a transformation that acts nontrivially only in the $N=1$ sector, taking \bea
|01 \rangle \langle 01|\to \frac{1}{2}(|01 \rangle \langle 01|+|10 \rangle \langle 10|),
\eea
and 
\bea
|10 \rangle \langle 10|\to \frac{1}{2}(|01 \rangle \langle 01|+|10 \rangle \langle 10|).
\eea
This is an example of classical communication creating classical correlations between $A$ and $B$ but no entanglement. The Lindblad operators describing this process are 
\be
\begin{split}
K_1=\frac{1}{\sqrt{2}}|10\rangle\langle 10|,\quad K_2=\frac{1}{\sqrt{2}}|01\rangle\langle 01|,\\
K_3=\frac{1}{\sqrt{2}}|10\rangle\langle 01|,\quad K_4=\frac{1}{\sqrt{2}}|01\rangle\langle 10|.
 \end{split}
\ee
These operators satisfy conditions (i), (ii) and (iii). Consider an initial density matrix 
\be
\rho=a|10\rangle\langle 10|+(1-a)|01\rangle\langle 01|+b|10\rangle\langle 01|+b^*|01\rangle\langle 10|,
\ee
whose NE $\Delta S_m(\rho)>0$ since it has a finite subsystem charge coherence $|b|\ne 0$. Now consider $\mathcal{K}(\rho)$,
\be
\mathcal{K}(\rho)=\sum_n K_n\rho K_n^\dagger= \frac{1}{2}|10\rangle\langle 10|+\frac{1}{2}|01\rangle\langle 01|.
\ee
In this case, $\Delta S_m(\mathcal{K}(\rho)) = 0$, which satisfies Eq.~\eqref{eq:monotonicity}.

\section{Relation between $\Delta S_m$ and negativity} 
\label{se:negativity}
\begin{figure}[b]
	\includegraphics[scale=0.6]{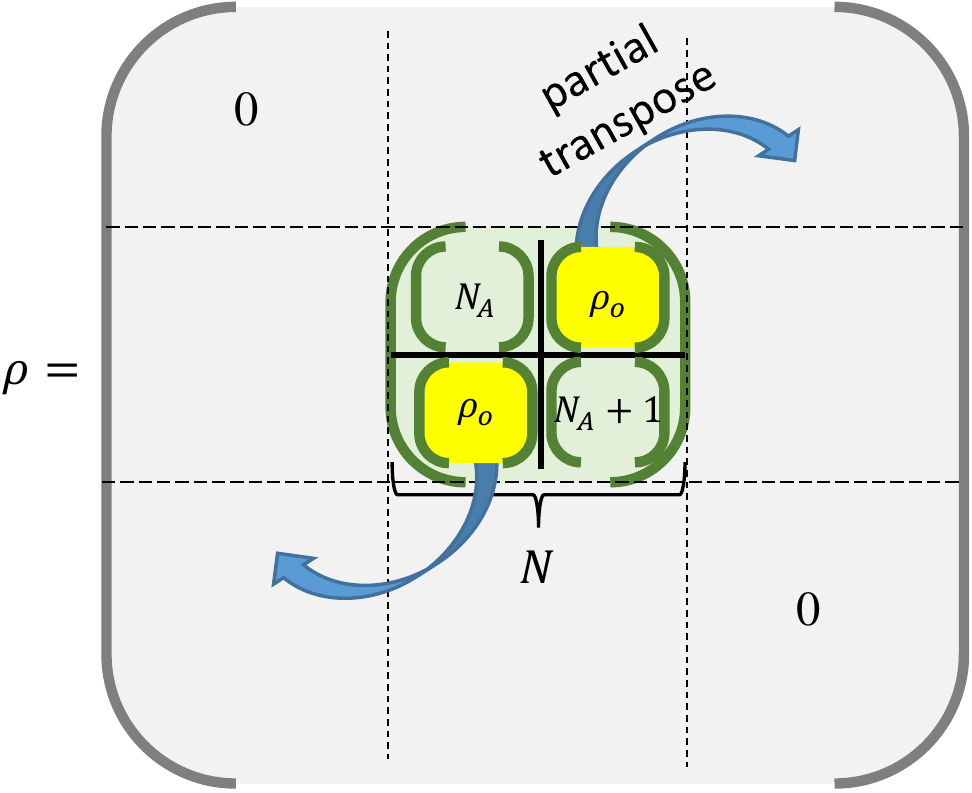}
	\caption{The total charge-$N$ block of the density matrix contains both diagonal and off diagonal subblocks in terms of the subsystems charge $N_A$. Under partial transposition  the off-diagonal blocks map to different total charge sectors, see Eq.~(\ref{rhooff}). Under unselective measurement $\rho \to \rho_m$, the off-diagonal blocks are annihilated.
	}
	\label{fig:01}
\end{figure}

Having demonstrated that  $\Delta S_m$ witnesses symmetric-inseparability, we now discuss its relation to negativity. Consider  mixed states with a fixed 
number of particles, $N$. Logarithmic negativity is defined by $\mathcal{N}=\log || \rho^{T_A} ||$ where $^{T_A}$ represents a partial transposition with respect to subsystem $A$. We make two statements:
\begin{enumerate}
	\item If subsystem $A$ has only one site, i.e. its state is fully specified by $N_A$, then the set of states with zero negativity equals the set of states with zero $\Delta S_m$. 
	\item If subsystem $A$ contains more than 1 site, the set of states with zero negativity is included in the set of states with zero $\Delta S_m$. 
\end{enumerate}
The second more general statement is illustrated in Fig.~\ref{fig:0}(a) and is proven as follows.
The most general mixed state with a fixed total particle number can be written as
\bea
\rho &=& \sum_{N_A,N_A' ,i,i',\alpha,\alpha'} C_{i,i',\alpha,\alpha'}^{N_A,N_A'} ~~~ \times  \\
&&|N_A , i \rangle_A \langle N_A',i'| \otimes   |N-N_A , \alpha \rangle_B \langle N- N_A',\alpha'|.\nonumber
\eea
By considering separately terms in the sum with $N_A=N_A'$ and $N_A \ne N_A'$, we split the density matrix into two parts  $\rho  = \rho_d+\rho_o$, being either diagonal or off-diagonal with respect to $N_A$, respectively, see Fig.~\ref{fig:01}. If we perform a partial transposition with respect to $A$ we obtain
\bea
\label{rhooff}
\rho_d^{T_A} &=& \sum_{N_A,i,i',\alpha,\alpha'} C_{i,i',\alpha,\alpha'}^{N_A,N_A}  \times \\ &&|N_A , i' \rangle_A \langle N_A,i| \otimes   |N-N_A , \alpha \rangle_B \langle N- N_A,\alpha'|, \nonumber \\
\rho_o^{T_A} &=& \sum_{N_A \ne N_A' ,i,i',\alpha,\alpha'} C_{i,i',\alpha,\alpha'}^{N_A,N_A'}  ~~~\times  \nonumber \\ &&|N_A' , i' \rangle_A \langle N_A,i| \otimes   |N-N_A , \alpha \rangle_B \langle N- N_A',\alpha'|.\nonumber
\eea
From these expressions it is clear that (i) $\rho_d^{T_A}$ still lies in the same symmetry sector with total particle number $N$. On the other hand $\rho_o^{T_A}$ lies completely outside of the original symmetry sector, see Fig.~(\ref{fig:01}); (ii)
Both $\rho_d^{T_A}$ and $\rho_o^{T_A}$ are hermitian matrices because partial tranposition preserves hermiticity; (iii) ${\rm{Tr}}\rho_o^{T_A}=0$. 

Because $\rho_d^{T_A}$ and $\rho_o^{T_A}$ lie in completely different sectors we can diagonalize them separately. The matrix $\rho_o^{T_A}$ is traceless and hermitian. Thus all of its eigenvalues are real and sum up to zero. Hence negative eigenvalues are guaranteed unless all the eigenvalues vanish, i.e., all the matrix elements vanish.

Therefore, zero negativity means at least that $\rho_o$ vanishes. The inverse is not true because some residue of the effect of partial tranposition acting on the internal $i$ degrees of freedom. However, if there are no other degrees of freedom besides the local particle number in subsystem $A$, e.g. if it contains only one site, then the set of states with zero negativity equals the set of states with no off-diagonal elements with respect to $N_A$.

Now consider mixed states with fluctuating total $N$ as in Fig.~\ref{fig:0}(b). From statement (2) above we  deduce  our main statement:  \emph{If a symmetric state has $\Delta S_m>0$ then it necessarily contains negativity in some charge sectors. }  Thus, entanglement can be extracted by projection to a fixed total charge sector.  This statement follows because $\Delta S_m>0$ ensures that there exists at least one charge-$N$ block with finite $\rho_o$.

\begin{figure}[b]
	\includegraphics[width=\linewidth]{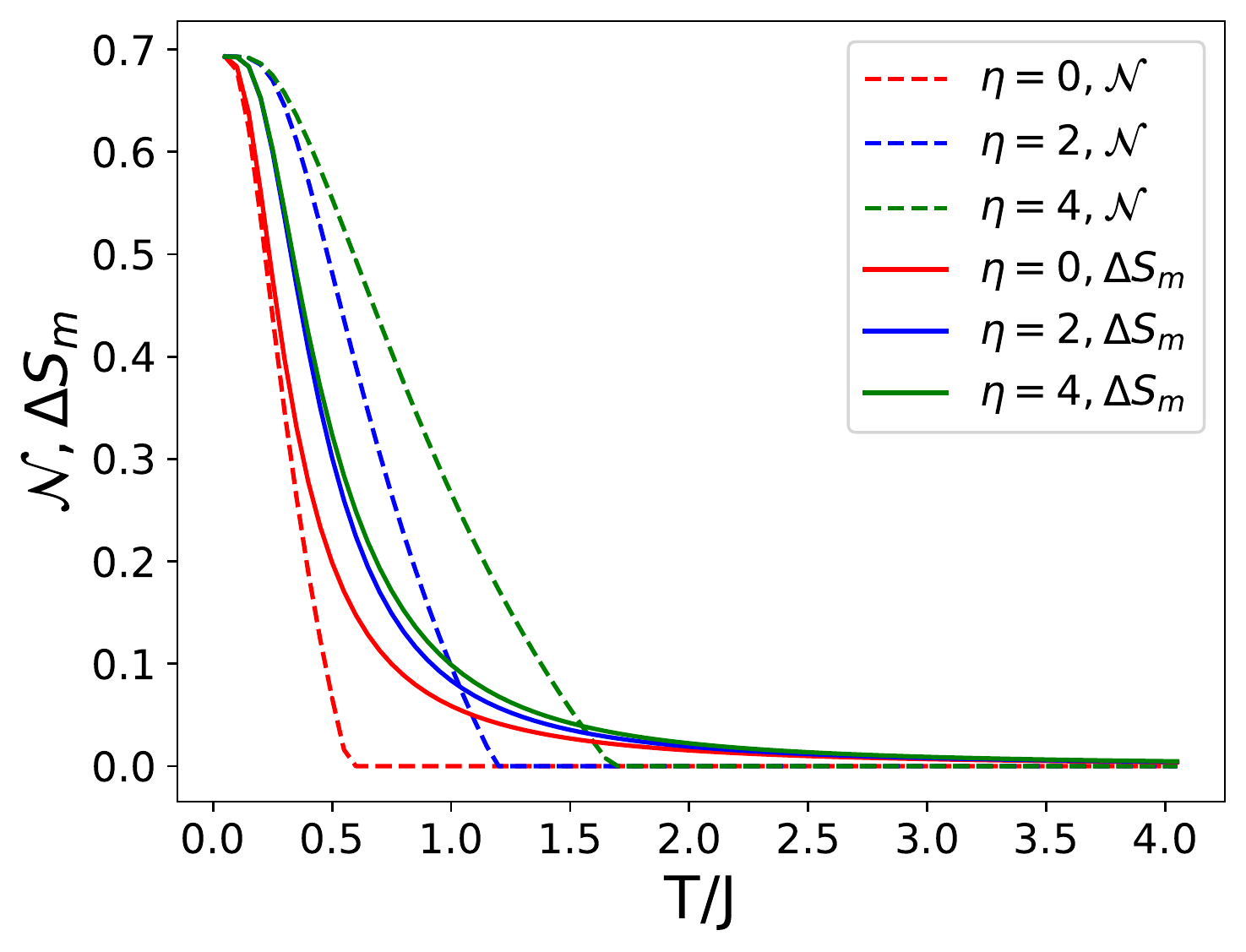}
	\caption{Entropy change $\Delta S_m$ and logarithmic negativity $\mathcal{N}$ for the two-site XXZ model Eq.~(\ref{eq:q:XXZmodel}). The leftmost, middle, rightmost solid (dashed) lines correspond to $\Delta S_m$ ($\mathcal{N}$) for $\eta=0, 2, 4$ respectively. While the latter displays a sudden death above some interaction dependent temperature~\cite{sherman2016nonzero}, we have $\Delta S_m>0$ indicating finite entanglement in fixed sectors at any temperature. 
	}
	\label{fig:2site}
\end{figure}

The generalization of the concept of negativity based on partial tranposition to fermionic systems had been a challenge addressed recently~\cite{shapourian2017partial,shapourian2019entanglement,Shapourian2019TwistedAU}.   In Appendix~\ref{app:fermionicnegativity} we comment on the comparison of $\Delta S_m$ and fermionic negativity.

\section{Examples}
\label{se:examples}
\subsection{Sudden death of entanglement in the XXZ model}
We now consider an example illustrating that $\Delta S_m>0$ implies entanglement in fixed charge sectors even when $\mathcal{N}=0$. As a standard interacting model with a  U(1) symmetry, consider the XXZ Hamiltonian 
\be
\label{eq:q:XXZmodel}
H_{XXZ}=J\sum_{i=1}^{L-1}( s_i^x s_{i+1}^x+ s_{i}^y s^y_{i+1}+\eta s_{i}^z s_{i+1}^z),
\ee 
where $s_i^a$ are spin $1/2$ operators acting on site $i$. This 1D model is equivalent to interacting  hard core bosons, or fermions, with hopping amplitude $t=\frac{J}{2}$ and  interaction $J\eta$. $s^z$ conservation maps to a particle number conservation. The thermal state $\rho =\frac{e^{-\beta H_{XXZ}}}{Z}$ was shown~\cite{sherman2016nonzero} to display a sudden death of negativity at some critical temperature. However we find that  $\Delta S_m>0$ at any temperature. The comparison of $\mathcal{N}$ and $\Delta S_m$ is plotted in Fig.~\ref{fig:2site} in the simple case  with $L=2$ and $L_A=1$. The thermal density matrix can be written as
\begin{align*}
\rho &=\frac{1}{Z} \bordermatrix{~ & \ket{00} & \ket{01} & \ket{10} & \ket{11} \cr
	&  e^{-\frac{\beta\eta}{4}} & 0 & 0 & 0 \cr
	& 0 & e^{\frac{\beta\eta}{4}}\cosh{\frac{\beta}{2}} & -e^{\frac{\beta\eta}{4}}\sinh{\frac{\beta}{2}} & 0 \cr
	& 0 &  -e^{\frac{\beta\eta}{4}}\sinh{\frac{\beta}{2}} & e^{\frac{\beta\eta}{4}}\cosh{\frac{\beta}{2}} & 0\cr
	& 0 & 0 & 0 &  e^{-\frac{\beta\eta}{4}}}.
\end{align*} 
We can see that there is entanglement in the $N=1$ sector, which is encoded in the off-diagonal elements $\propto \sinh{(\beta/2)}$. 
A similar situation occurs for thermal free bosons as discussed in  Appendix~\ref{app:thermalbosons}.
Extending on this example, we now construct examples of separable states with $\Delta S_m>0$, either for bosons or fermions.

\subsection{Separable states with $\Delta S_m >0$}
We now provide two examples of separable states with $\Delta S_m >0$, as marked with diagonal lines in Fig.~\ref{fig:0}(b). Although these states are separable they contain entanglement in specific charge sectors. The only way to achieve a separable form, is by violating Eq.~(\ref{symmetrycondition}) on the level of each classically combined component.

\subsubsection{2-site spin state}
As an example illustrating that $\Delta S_m>0$ implies entanglement in fixed charge sectors even for separable states, consider the two site example 
\bea
\label{eq:qD1}
\rho &=&\frac{1}{4}(|x_+ \rangle \langle x_+| \otimes |x_- \rangle \langle x_-|+ |x_- \rangle \langle x_-| \otimes |x_+ \rangle \langle x_+| \nonumber \\
&+&|y_+ \rangle \langle y_+| \otimes |y_- \rangle \langle y_-|+ |y_- \rangle \langle y_-| \otimes |y_+ \rangle \langle y_+|),
\eea
where $|x_\pm \rangle$ and $|y_\pm \rangle$ are states with $\pm 1$ eigenvalues of the Pauli-matrix operators $\sigma^x$ and $\sigma^y$, respectively.  Mapping spin $\uparrow(\downarrow)$ to an occupied (empty) site, we use the basis $\{ | 00 \rangle , | 01 \rangle, | 10 \rangle, | 11 \rangle \}$. This state  can be written as
$\rho=\frac{1}{4}\left( |00 \rangle \langle 00 |+|11 \rangle \langle 11 |  \right) + \frac{1}{2} | \psi_- \rangle \langle \psi_- |$, where $|\psi_-\rangle = \frac{|01 \rangle - |10 \rangle}{\sqrt{2}}$. We see that this explicitly separable state  has a block structure, i.e. $[\rho,\hat{N}]=0$. Clearly, this state has $\Delta S_m>0$ and it indeed has a nonseparable charge sector of $N=1$. Neither of the components of Eq.~(\ref{eq:qD1}) conserves the symmetry, only their sum does.
\subsubsection{Fermions}
\label{se:fermionsexample}
In dealing with fermionic systems we first deal with parity conservation. Separable states are defined as in Eq.~(\ref{eq:separability}) where, importantly, we require
\be
[\rho^i_A, (-1)^{N_A}]=0.
\ee
We devide non-separable states into 2 branches~\cite{shapourian2019entanglement}:
\begin{enumerate}
	\item  States with a block diagonal form in terms of the fermion-number parity of the subsystem,
	\be
	[\rho, (-1)^{F_A}]=0.
	\ee
	\item States containing off-diagonal blocks in terms of the fermion-number parity of the subsystem,
	\be
	[\rho, (-1)^{F_A}]\neq 0.
	\ee
\end{enumerate}
It is clear that $\Delta S_m^{({\rm{parity}})}$, the entropy change induced by a parity measurement, is non-zero for all non-separable states in branch 2. 

Now, suppose that besides the fermion number parity, we have an additional $U(1)$ symmetry. As in the bosonic case, we could find separable states where each decomposition does not conserve this $U(1)$ symmetry.

Analogous to the bosonic counterexample, consider the following mixed state of 4 fermions
\begin{align*}
\rho &= \bordermatrix{~ & \ket{0000} & \ket{0011} & \ket{1100} & \ket{1111} \cr
	&  \frac{1}{4} & 0 & 0 & 0 \cr
	& 0 & \frac{1}{4} & -\frac{1}{4} & 0 \cr
	& 0 &  -\frac{1}{4} & \frac{1}{4} & 0\cr
	& 0 & 0 & 0 &  \frac{1}{4}},
\end{align*} 
where the subsystem $A$ consists of the first 2 sites. This mixed state has a decomposition similar to the mixed state of 2 qubits, while preserving the local fermion  parity. This separable state has zero negativity, but has a finite $\Delta S_m$. This state is of branch 1.

\begin{figure*}[t]
	\includegraphics[width=\linewidth]{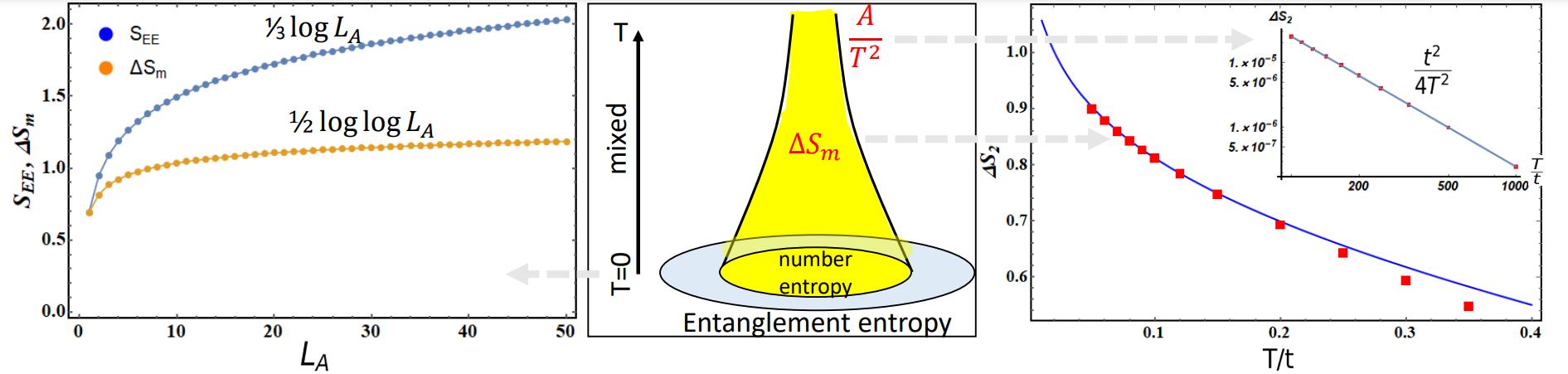}
	\caption{Middle panel: schematic temperature dependence of $\Delta S_m$ for general systems. At $T \to 0$ it approaches the number entropy, which is contained in the entanglement entropy $S_{EE}$. At high temperature it decays as $1/T^2$ following an area law. Left panel: comparison of $S_{EE}$ (upper) and $\Delta S_m$ (lower) at $T=0$ for free fermions and $L \to \infty$, $L_A$ denotes the number of sites in subsystem $A$. Analytic fits are $\Delta S_m \cong \frac{1}{2} \log ( 1.731 (\log L_A+2.269))$ and $S_{EE} \cong \frac{1}{3} \log L_A+.726$. Right panel: Second \Renyi entropy $\Delta S_2$ for $L=1000$ and $L_A=100$ using numerical (squares) and CFT results. 
		Inset: fit to $1/T^2$ form.
	}
	\label{fig:comb}
\end{figure*}

\section{Calculation methods and scaling properties} 
\label{se:methods}

We are now interested in the scaling of $\Delta S_m$ with $T$, $L$ and $L_A$. The von Neumann entropy $\Delta S_m$ can in principle be calculated from its \Renyi moments via the replica trick using analytic methods~\cite{calabrese2004entanglement}. Leaving this formidable task to future work, here we focus on the second moment of the NE, i.e. the change of second \Renyi entropy $\Delta S_2 = S_2(\rho_m) -S_2(\rho)$, where $S_2=-\log{\Tr{\rho^2}}$.

According to Eq.~(\ref{eq:14}), for a thermal state $\rho = \frac{e^{- \beta H}}{Z}$, we need to calculate
\be
\label{eq:trace2beta}
\Tr{\rho_m^2}=\int\frac{d\alpha_1d\alpha_2}{(2\pi)^2}\frac{1}{Z^2}\Tr(e^{-\beta H}e^{-i \alpha_{12} \hat{N}_A}e^{-\beta H}e^{i\alpha_{12} \hat{N}_A}),
\ee
where $\alpha_{12}=\alpha_1-\alpha_2$.

We apply this formula as a starting point for various methods: (i) numerical calculation for free fermions, (ii) conformal field theory (CFT), (iii) high temperature expansion. To illustrate these methods below, the model of interest is a free fermion  chain, $H=-t\sum_i(c_{i+1}^{\dag}c_i+h.c.)$.

\subsection{Numerical results}
We first develop a numerical method to calculate $\Delta S_2$ (and similar quantities) in lattice models of free fermions. The method is based on properties of Gaussian operators. Specifically\par
\be
\label{eq:gaussianproduct}
e^{c_i^{\dag}A_{ij}c_j}e^{c_m^{\dag}B_{mn}c_n}=e^{c_k^{\dag}F_{kl}c_l},
\ee
where $F=\log{(e^Ae^B)}$. This can be proven using the Baker-Campbell-Hausdorff formula. Secondly\par
\be
\label{eq:gaussiantrace}
\Tr{e^{c_i^{\dag}S_{ij}c_j}}=\det(\mathcal{I}+e^S).
\ee
This holds true for a general non-Hermitian $S$, see Ref.~\onlinecite{fagotti2010entanglement}.
We start with Eq.~(\ref{eq:trace2beta}).
It is important to notice that both $e^{-\beta H}$ and $e^{-i(\alpha_1-\alpha_2) \hat{N}_A}$ are Gaussian operators. According to Eq.~(\ref{eq:gaussianproduct}), the product of 4 Gaussian operators is still Gaussian, and the trace can be calculated using Eq.~(\ref{eq:gaussiantrace}).\par
So suppose $H=\sum_{i,j}c^{\dag}_i h_{ij} c_j$, and $\hat{N}_A=\sum_{i,j}c^{\dag}_i n^A_{ij} c_j$. Then we conclude
\begin{multline}
\label{eq:singleparticle}
\Tr(e^{-\beta H}e^{-i(\alpha_1-\alpha_2) \hat{N}_A}e^{-\beta H}e^{i(\alpha_1-\alpha_2) \hat{N}_A}\\
=\det(\mathcal{I}+e^{-\beta h}e^{-i\alpha_{12}n^A}e^{-\beta h}e^{i\alpha_{12}n^A}).
\end{multline}
The calculation of $Z$ in Eq.~(\ref{eq:trace2beta}) and the second \Renyi entropy of the un-measured state $S_2(\rho)$  is straightforward. 

Using the above methods, we calculate $\Delta S_2$ in a chain of size $L=1000$, the subsystem size is fixed to be $L_A=100$. We plot the temperature dependence of $\Delta S_2$ in the right panel of Fig.~\ref{fig:comb}, as square symbols.

\subsection{High temperature limit}

At temperatures $T \gg t$, performing a high temperature expansion we find that in a general lattice with hopping amplitude $t$ and area $\mathcal{A}$ separating $A$ and $B$, the NE becomes
\be
\Delta S_m \to (t^2\mathcal{A})/(4T^2)
\ee
The derivation, which is essentially an expansion in $\beta$ of Eq.~(\ref{eq:trace2beta}), is given in Appendix~\ref{Appendic:HighT}.  This form holds true for the second \Renyi entropy as shown in the inset of the right panel of Fig.~\ref{fig:comb}.

\subsection{CFT methods}
To obtain the NE at $T=0$, we borrow results for the number entropy~\cite{calabrese2021symmetry,goldstein2018symmetry} $\Delta S_m \sim \frac{1}{2}  \log \left(\frac{2K}{\pi} \log L_A \right)$ where $K$ is the Luttinger parameter. This is demonstrated in the left panel of Fig.~\ref{fig:comb} for noninteracting fermions, and also compared with the entanglement entropy $S_{EE}$~\cite{calabrese2004entanglement}. While the two coincide for a single site in $A$, $S_{EE} \sim \frac{1}{3} \log L_A$ exceeds $\Delta S_m$, since the NE only captures entanglement between charge degrees of freedom.

We now are interested in the scaling of the NE at finite temperature, which are either small or large compared to the level spacing within the subsystem ($\propto 1/L_A$). Using CFT methods, we can express $\Delta S_2 = - \log \frac{{\rm{Tr}} \rho_m^2}{{\rm{Tr}} \rho^2}=-\log{\int\frac{d\alpha_1d\alpha_2}{(2\pi)^2}\braket{V_1V_2V_3V_4}}_{2 \beta}$ in terms of a correlation function on a cylinder of circumference $2 \beta$, with $\braket{V_1V_2V_3V_4}_{2 \beta}=(\frac{2\beta}{\pi}\tanh{\frac{\pi L_A}{2\beta}})^{-\frac{(\alpha_1-\alpha_2)^2}{\pi^2}}$. The details of the derivation are given in Appendix~\ref{Appendic:cft}. This is valid for any ratio $\beta / L_A$, as long as $T$ is lower than the high energy cutoff $\Lambda$, set by $t$. From this, we derive a crossover from the low temperature regime discussed above, to an intermediate temperature regime, $\frac{1}{L_A} \ll T \ll \Lambda$, where we find $\Delta S_2 \sim \frac{1}{2} \log \log \beta$. This is compared with numerical results in the right panel of Fig.~\ref{fig:comb}. Deviations from CFT results are seen for $T \lesssim \Lambda$.

The result for the NE combining these various methods is summarized in the central panel of Fig.~\ref{fig:comb}. We can see that at zero temperature the NE coincides with the number entropy, which itself is only a part of the entanglement entropy - except when the subsystem has only one site.  The definition of the NE $\Delta S_m$ extends to finite temperature, where at very high temperature it decays as $1/T^2$ according to an area law.

\section{Summary} 
\label{se:summary}
A projective measurement whose result is not being recorded increases the entropy of a quantum system. When applied to a subsystem in a conserved charge basis, it annihilates any coherences between blocks with different subsystem charges. The entropy change, referred to as number entanglement (NE), indicates entanglement, or inseparability between the measured subsystem and its complement. 

The NE quantifies entanglement in mixed states as long as the density matrix commutes with the symmetry. It goes beyond other quantities such as entanglement entropy which is restricted to pure states, or logarithmic negavity which does not account for the symmetry resolution of inseparability.

There are a number of directions to measure $\Delta S_m$ in experiment. These include cold atom experiments realizing the replica trick~\cite{islam2015measuring}, which also allow to measure negativity~\cite{gray2017measuring,cornfeld2018imbalance}, as well as experiments realizing random unitaries~\cite{elben2019statistical,vitale2021symmetry}. Another promising direction is based on mesoscopic systems. Recently, it was demonstrated how to measure changes of entropy~\cite{hartman2018direct,kleeorin2019HowTM,sela2019detecting}. We envision $\Delta S_m$ as a special case of an entropy change occurring as we turn on a nearby mesoscopic conductor acting as charge detector, which causes dephasing and decoherence~\cite{aleiner1997dephasing,levinson1997dephasing}.

	\section{Acknowledgments} We acknowledge support from  the European Research Council (ERC) under the European Unions Horizon 2020 research and innovation programme under grant agreement
	No. 951541, ARO (W911NF-20-1-0013), the US-Israel Binational
	Science Foundation (Grant No. 2016255) and the Israel Science Foundation, grant number 154/19. We thank discussions with Pasquale Calabrese, Marcello Dalmonte, Moshe Goldstein, and Tarun Grover.

\appendix

\section{Relation to symmetry-resolved entanglement}
\label{app:sre}
In this appendix, we discuss the entropy change upon measurement, $\Delta S_m$, in the context of entanglement entropy and symmetry resolved entanglement~(SRE). 

For pure states with a global conserved charge the entanglement entropy can be separated as
\be
\label{eq:purestatedec}
S(\rho_A)=H_1(\{ P(N_A) \})+\sum_{N_A} P(N_A) S(\rho_A(N_A)),~~~~~~(T=0).
\ee
Here $H_1(\{ P(N_A) \}) = -\sum_{N_A} P(N_A) \log P(N_A)$ is the Shannon entropy of the subsystem charge
probability distribution, which we refer to as number entropy. Also it is often referred to as inaccessible entanglement~\cite{barghathi2018renyi,barghathi2019operationally}. The second term~\cite{wiseman2003entanglement} 
is the weighted contribution of the SRE originating from each superselection sector corresponding to $N_A$ particles in $A$, where 
\be
\rho_A(N_A)=\frac{1}{P(N_A)} \Pi(N_A) \rho_A \Pi(N_A).
\label{eq:reducedmeasure}
\ee
This separation of the EE into a number entropy and the weighted SRE is displayed in Fig.~(\ref{fig:comb}) of the main text. Relatedly, connections between entanglement entropy and charge fluctuations were emphasized in 1D~\cite{song2010general,song2011entanglement,rachel2012detecting}. 

In this work we consider a general mixed density matrix $\rho$ and our quantity of interest is
\be
\Delta S_m = H_1(\{ P(N_A) \})+\sum_{N_A} P(N_A)[S(\rho(N_A)) - S(\rho)].
\label{eq:deltasm1}
\ee
Here, compared to the pure state decomposition Eq.~(\ref{eq:purestatedec}), in the second term we have the entropy of the full state acting both on $A$ and $B$ after it has been projected and normalized to a given number of particles in $A$,
\be
\rho(N_A)=\frac{1}{P(N_A)} \Pi(N_A) \rho \Pi(N_A).
\ee
Note the difference compared to Eq.~(\ref{eq:reducedmeasure}) which involves the reduced density matrix. Thus the second term in Eq.~(\ref{eq:deltasm1}) is the weighted entropy change for each charge state. Interestingly, this is equivalent to our simple definition Eq.~(\ref{eq:maindef}).
Note that in Eq.~(\ref{eq:maindef})  $\rho_m = \sum_{N_A} \Pi(N_A) \rho \Pi(N_A)= \oplus_{N_A} P(N_A) \rho(N_A)$ is normalized but consists of non-normalized blocks.

\section{Thermal state of two boson modes}
\label{app:thermalbosons}
\begin{figure}[h]
	\includegraphics[width=\linewidth]{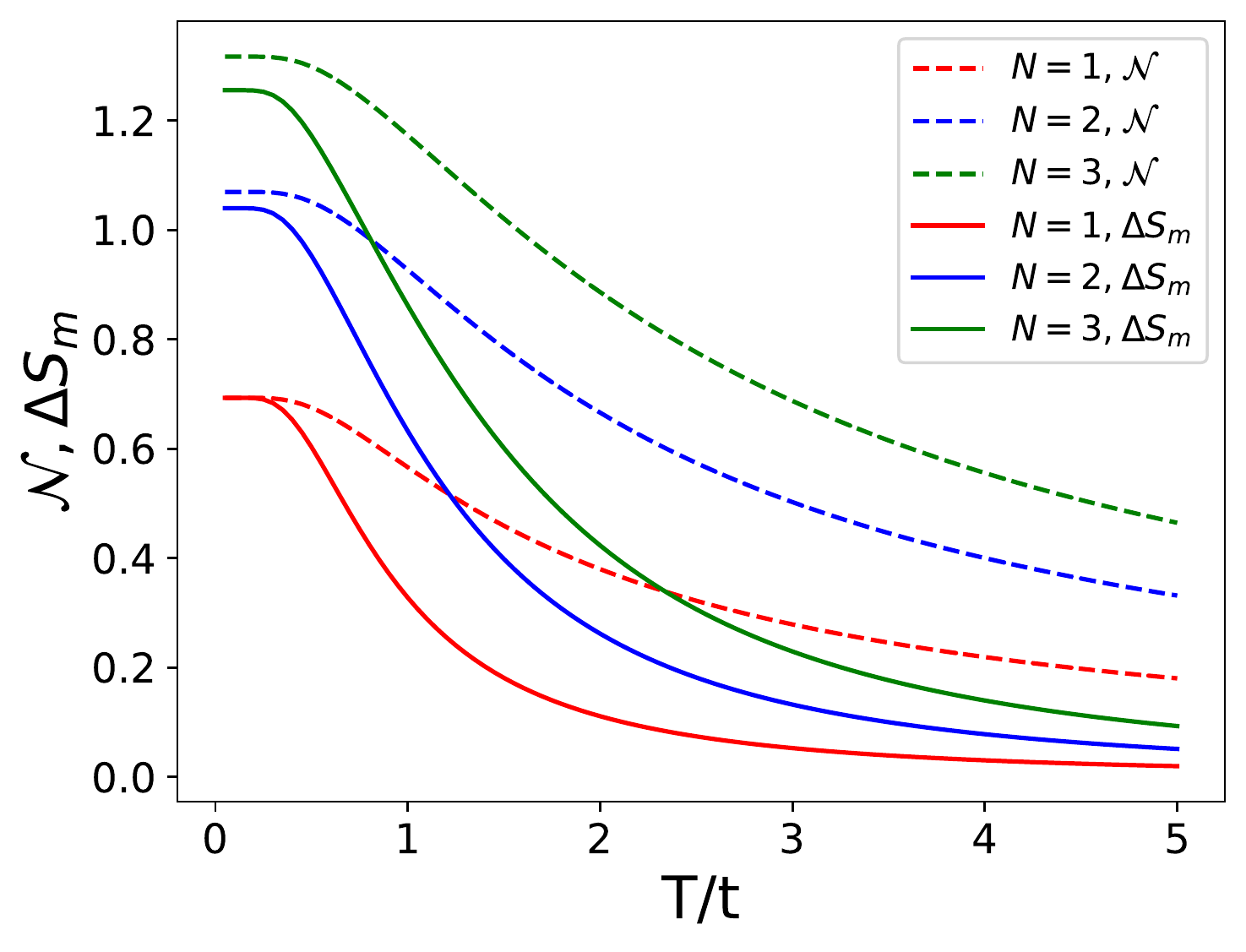}
	\caption{The uppermost, middle, lowermost solid (dashed) lines correspond to $\Delta S_m$ ($\mathcal{N}$) for $N=3, 2, 1$ respectively.
	}
	\label{fig:thermalboson}
\end{figure}
We consider the Hamiltonian
\par
\begin{align*}
H &= \begin{pmatrix}
a_1^{\dag} & a_2^{\dag}
\end{pmatrix}
\begin{pmatrix}
-\mu & t\\
t   & -\mu
\end{pmatrix}
\begin{pmatrix}
a_1\\
a_2
\end{pmatrix},
\end{align*}
which conserves the total particle number $N$. The Gibbs state $\rho = e^{-\beta H}$ was shown to be separable~\cite{lu2020structure}. Hence the negativity of the full state vanishes. 

However, according to our results, some of the symmetry sectors $\rho(N) = \Pi(N) \rho \Pi(N)$ are entangled, because the off-diagonal terms in $N_A$ generally exist in the occupation number basis. In the following, we calculate $\Delta S_m$ and negativity in the $N=1,2,3$ sectors, and  plot both of them in Fig.~(\ref{fig:thermalboson}).

The state we consider is $\rho = \frac{1}{Z} e^{-\beta H} \Pi(N)$, or  equivalently  $\rho = \frac{1}{Z} e^{-\beta H(N)}$ where $H(N)$ is the charge-$N$ block of $H$.

In the 1 boson subspace, $H$ can be written as
\begin{align*}
H_1 &= \bordermatrix{~ & \ket{10} & \ket{01} \cr
	&  -\mu & t \cr
	& t & -\mu \cr}.
\end{align*} 
Thus we have
\begin{align*}
\rho=\frac{1}{Z}e^{-\beta H_1} &= \begin{pmatrix}
\frac{1}{2} & -\frac{1}{2}\tanh{\beta}\\
-\frac{1}{2}\tanh{\beta} & \frac{1}{2}
\end{pmatrix}.
\end{align*}

After simple algebra, we find  that the negativity and the entropy change are
\begin{align}
\mathcal{N}&=\log{(1+\tanh{\beta})}, \\
\Delta S_m &= \log{2} + \frac{1-\tanh{\beta}}{2}\log{\frac{1-\tanh{\beta}}{2}}\nonumber\\
&\qquad\qquad + \frac{1+\tanh{\beta}}{2}\log{\frac{1+\tanh{\beta}}{2}}.
\end{align}
Similarly we could derive the negativity and entropy change in $N=2,3$ subspaces, as plotted in Fig.~(\ref{fig:thermalboson}).\\

\section{Fermionic negativity}
\label{app:fermionicnegativity}
In this appendix we comment on the comparison of $\Delta S_m$ and fermionic negativity.
The generalization of the concept of negativity based on partial tranposition to fermionic systems had been a challenge addressed recently~\cite{shapourian2017partial,shapourian2019entanglement,Shapourian2019TwistedAU}. First, the example in Sec.~\ref{se:fermionsexample} gives a fermionic state with $\Delta S_m$, but zero fermionic negativity. In this sense, also the fermionic negativity fails to account for the additional $U(1)$ symmetry that can allow to extract entanglement by projecting to specific sectors.

Below we show results for the fermionic negativity $\mathcal{N}^f$ and $\Delta S_m$ for the same XXZ 2-site system studied in the main text. We first provide definitions.

To define fermionic negativity, we first need to define fermionic partial transpose. In the occupation number basis,
\bea
\ket{\{n_j\}_{j\in A}, \{n_j\}_{j\in B} }&=&  \nonumber \\
(f_{j_1}^{\dag})^{n_{j_1}}\cdots(f_{j_{m_A}}^{\dag})^{n_{j_{m_A}}}&\cdots&(f_{j^{'}_{m_B}}^{\dag})^{n_{j^{'}_{m_B}}}\ket{0},
\eea
the fermionic partial transpose is defined as
\bea
&&(\ket{\{n_j\}_{j\in A}, \{n_j\}_{j\in B} }\bra{\{\bar{n}_j\}_{j\in A}, \{\bar{n}_j\}_{j\in B} })^{R_A}=\\
(-&1&)^{\phi{({n_j},{\bar{n}_j})}}U_A^{\dag}\ket{\{\bar{n}_j\}_{j\in A}, \{n_j\}_{j\in B} }\bra{\{n_j\}_{j\in A}, \{\bar{n}_j\}_{j\in B} }U_A. \nonumber 
\eea
To make a distinction between fermionic partial transpose and normal partial transpose, here we denote it as $(\ket{\cdots}\bra{\cdots})^{R_A}$.
We can see it is the same as normal partial transpose up to a phase factor. 

Because of this phase factor, the fermionic partial transpose of a density matrix is no longer Hermitian. The fermionic negativity is defined as
\be
\mathcal{N}^f = \ln{\Tr{\sqrt{\rho^{R_A}(\rho^{R_A})^{\dag}}}}.
\label{eq:fermionicnegativity}
\ee
As an explicit example, let's calculate the fermionic negativity in the XXZ model.

\begin{figure}[h]
	\includegraphics[width=\linewidth]{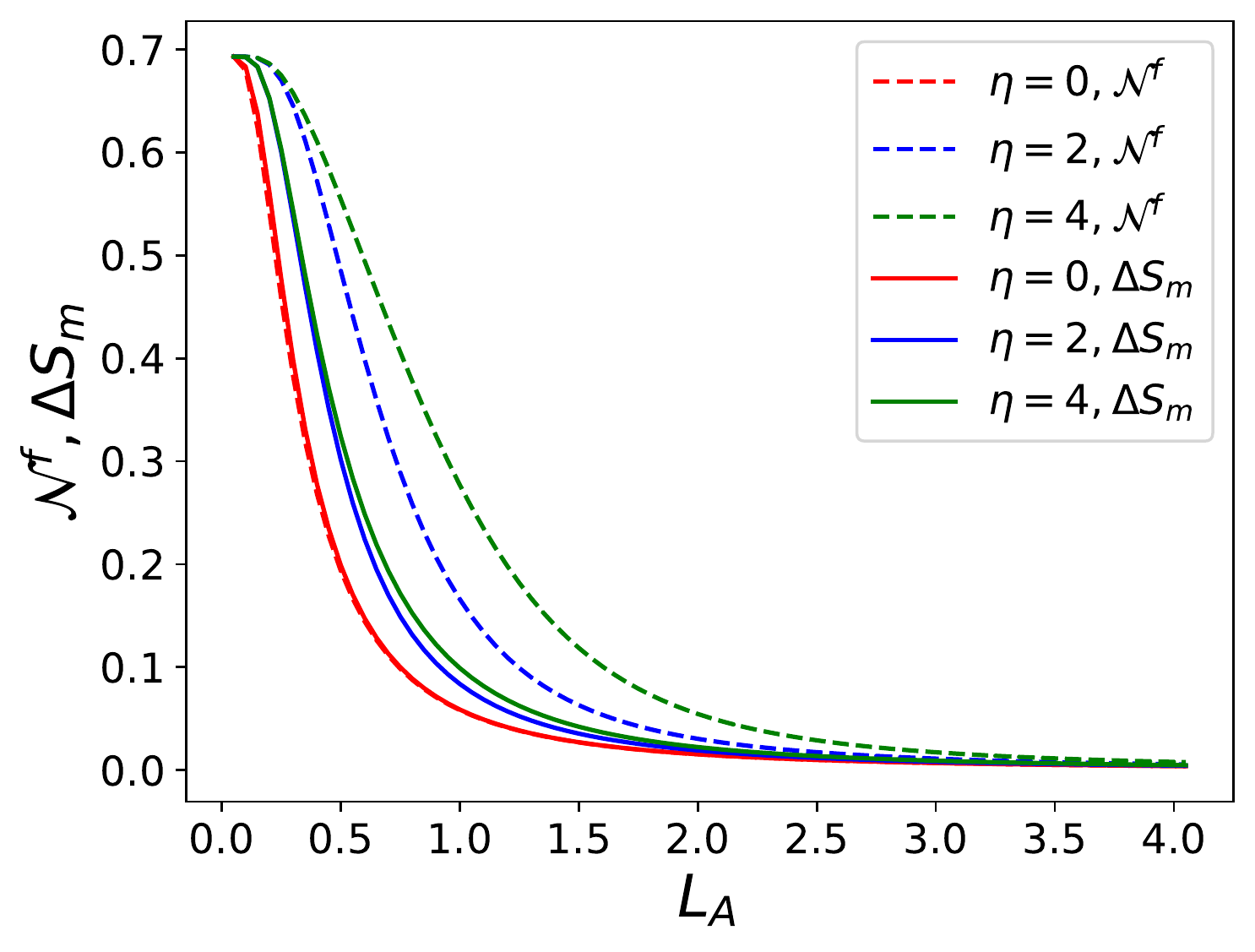}
	\caption{$\mathcal{N}^f$ and $\Delta S_m$ of $\rho$ in Eq.~(\ref{eq:fermidensity}), the leftmost, middle, rightmost solid (dashed) lines correspond to $\Delta S_m$ ($\mathcal{N}^f$) for $\eta=0, 2, 4$ respectively.}
	\label{fig:fermivsench}
\end{figure}

Performing Jordan-Wigner transformation to the 2-site XXZ Hamiltonian Eq.~(\ref{eq:q:XXZmodel}), we obtain
\be
H=J\left(-\frac{1}{2}(c_1c^{\dag}_2+c_2c^{\dag}_1) + \eta(1-2c_1^{\dag}c_1)(1-2c_2^{\dag}c_2)\right).
\ee
The thermal density matrix $\rho=e^{-\beta H}/Z$ is 
\begin{align}
\rho &=\frac{1}{Z} \bordermatrix{~ & \ket{00} & \ket{01} & \ket{10} & \ket{11} \cr
	&  e^{-\frac{\beta\eta}{4}} & 0 & 0 & 0 \cr
	& 0 & e^{\frac{\beta\eta}{4}}\cosh{\frac{\beta}{2}} & -e^{\frac{\beta\eta}{4}}\sinh{\frac{\beta}{2}} & 0 \cr
	& 0 &  -e^{\frac{\beta\eta}{4}}\sinh{\frac{\beta}{2}} & e^{\frac{\beta\eta}{4}}\cosh{\frac{\beta}{2}} & 0\cr
	& 0 & 0 & 0 &  e^{-\frac{\beta\eta}{4}}}.
\label{eq:fermidensity}
\end{align} 
The fermionic partial transpose of $\rho^{R_A} $ times $Z$ is
\begin{align*}
& \bordermatrix{~ & \ket{00} & \ket{01} & \ket{10} & \ket{11} \cr
	&  e^{-\frac{\beta\eta}{4}} & 0 & 0 & -i e^{\frac{\beta\eta}{4}}\sinh{\frac{\beta}{2}} \cr
	& 0 & e^{\frac{\beta\eta}{4}}\cosh{\frac{\beta}{2}} & 0 & 0 \cr
	& 0 & 0 & e^{\frac{\beta\eta}{4}}\cosh{\frac{\beta}{2}} & 0\cr
	& -i e^{\frac{\beta\eta}{4}}\sinh{\frac{\beta}{2}} & 0 & 0 &  e^{-\frac{\beta\eta}{4}}}.
\end{align*} 
Then we could calculate $\mathcal{N}^f$, as in Eq.~(\ref{eq:fermionicnegativity}). We plot the temperature and interaction strength dependence of $\mathcal{N}^f$ in Fig.~(\ref{fig:fermivsench})

\section{High temperature expansion}
\label{Appendic:HighT}
In this appendix, we work out the high-temperature expansion of $\Delta S_2$ and $\Delta S_m$ in the tight-binding chain of free fermions 
\be
\label{eq:tighth}
H=-t\sum_i(c_{i+1}^{\dag}c_i+h.c.).
\ee
The  analysis below shows that it is sensitive to the hopping terms in the Hamiltonian, and immune to the interactions, as one might expect. 
\par
According to Eq.~(\ref{eq:trace2beta}),
\be
\Delta S_2=-\log{\int\frac{d\alpha_1d\alpha_2}{(2\pi)^2}\frac{\Tr(e^{-\beta H}e^{-i\alpha_{12} \hat{N}_A}e^{-\beta H}e^{i\alpha_{12} \hat{N}_A})}{\Tr{e^{-2\beta H}}}}.
\ee
Here and in the following, $\alpha_{ij}=\alpha_i-\alpha_j$.
For the denominator, we have the high-temperature expansion,
\begin{multline}
\label{eq:highde}
\Tr{e^{-2\beta H}}\approx\Tr{(\mathcal{I}-2\beta H+\frac{(2\beta)^2}{2}H^2)}\\
=D_H-2\beta\Tr{H}+\frac{(2\beta)^2}{2}\Tr{H^2}.
\end{multline}
Clearly, $D_H$ denotes the dimension of the Hilbert space.

For the numerator, similarly we have
\begin{equation}
\label{eq:highnum}
\begin{aligned}
&\Tr{(e^{-\beta H}e^{-i\alpha_{12} \hat{N}_A}e^{-\beta H}e^{i\alpha_{12} \hat{N}_A})}\approx\\
&D_H-2\beta\Tr{H}+\beta^2\Tr{H^2}+\\
&\qquad\qquad\beta^2\Tr{(He^{-i\alpha_{12} \hat{N}_A}He^{i\alpha_{12} \hat{N}_A})}.
\end{aligned}
\end{equation}
In the above expression, we used the cyclic property of the trace.
Comparing Eq.~(\ref{eq:highde}) and Eq.~(\ref{eq:highnum}), we see that
\begin{align}
\label{eq:simple}
&\frac{\Tr(e^{-\beta H}e^{-i\alpha_{12} \hat{N}_A}e^{-\beta H}e^{i\alpha_{12}\hat{N}_A})}{\Tr{e^{-2\beta H}}}\nonumber\\
\approx&1+\beta^2\frac{\Tr{(He^{-i\alpha_{12} \hat{N}_A}He^{i\alpha_{12} \hat{N}_A})}-\Tr{H^2}}{D_H}.
\end{align}
From the above expression, we can infer that the lowest order of $\beta$ expansion of $\Delta S_2$ is the $\beta^2$ order. Eq.~(\ref{eq:simple}) can be simplified further,
\be
\Tr{(He^{-i\alpha_{12} \hat{N}_A}He^{i\alpha_{12} \hat{N}_A})}-\Tr{H^2}=\Tr{(HO)},
\ee
where
\be
O=e^{-i\alpha_{12}\hat{N}_A}He^{i\alpha_{12}\hat{N}_A}-H.
\ee
Most of the terms in $H$ actually commute with $\hat{N}_A$, except for those terms that live at the boundary.

Specifically, suppose,
\be
\hat{N}_A=c_1^{\dag}c_1+\cdots+c_m^{\dag}c_m,
\ee 
and that $H$ is given by Eq.~(\ref{eq:tighth}). Here $m=L_A$ is the number of sites of subregion $A$. Then the only terms which have a non-vanishing commutator with $\hat{N}_A$, thus contribute to $O$, are $H_{hop}=-t(c_0^{\dag}c_1+c_1^{\dag}c_0+c_m^{\dag}c_{m+1}+c^{\dag}_{m+1}c_m)$, which generate hopping of particles between $A$ and its complement.

The operator $O$ can be calculated using Baker-Campbell-Hausdorf formula,
\be
\label{eq:bchf}
\begin{aligned}
	&e^{-i\alpha_{12} \hat{N}_A}He^{i\alpha_{12}\hat{N}_A}\\
	=&H+i\alpha_{21}[\hat{N}_A,H]+\frac{(i\alpha_{21})^2}{2}[\hat{N}_A,[\hat{N}_A,H]]+\cdots
\end{aligned}
\ee
Interestingly, we find 
\begin{multline}
[\hat{N}_A, H_{hop}]=
\\-t(-c_0^{\dag}c_1+c_1^{\dag}c_0+c_m^{\dag}c_{m+1}-c^{\dag}_{m+1}c_m),
\end{multline}
\begin{multline}
[\hat{N}_A,[\hat{N}_A, H_{hop}]]=
\\-t(c_0^{\dag}c_1+c_1^{\dag}c_0+c_m^{\dag}c_{m+1}+c^{\dag}_{m+1}c_m)=H_{hop}.
\end{multline}
Then Eq.~(\ref{eq:bchf}) yields
\begin{multline}
e^{-i\alpha_{12}\hat{N}_A}H_{hop}e^{i\alpha_{12} \hat{N}_A}=\\
\cos{\alpha_{21}}H_{hop}+i \sin{\alpha_{21}}[\hat{N}_A, H_{hop}].
\end{multline}
So we conclude
\be
O=\cos{\alpha_{21}}H_{hop}+i \sin{\alpha_{21}}[\hat{N}_A, H_{hop}]-H_{hop}.
\ee
To calculate $\Tr{(HO)}$, notice that
\begin{align}
\Tr{c_i^{\dag}c_j}&=\frac{1}{2}\delta_{ij}D_H,\\
\Tr{c_i^{\dag}c_j c_m^{\dag}c_n}&=\frac{1}{4}(\delta_{ij}\delta_{mn}+\delta_{in}\delta_{mj})D_H.
\end{align}
The only part in $H$ which contributes to $\Tr{(HO)}$ is $H_{hop}$ again.\par
Then
\begin{multline}
\Tr{(HO)}=(\cos{\alpha_{21}}-1)\Tr{H_{hop}^2}+\\
i \sin{\alpha_{21}}\Tr{(H_{hop}[\hat{N}_A, H_{hop}])}.
\end{multline}
After integration over $\alpha_2$ and $\alpha_1$, we conclude
\be
\Delta S_2=-\log{(1-\beta^2\frac{\Tr{H_{hop}^2}}{D_H})}\approx\beta^2\frac{\Tr{H_{hop}^2}}{D_H}=\beta^2t^2.
\ee
This result holds when $\beta^2t^2\ll 1$.

For a general lattice in any dimension, with nearest neighbor hopping, we have
\be
\Delta S_2 = \frac{\mathcal{A}t^2}{2T^2},
\ee
where the hopping amplitude $t$ is assumed to be constant and $\mathcal{A}$ is the area between $A$ and $B$. For our 1D case, we have $\mathcal{A}=2$.\par
Now consider a fermion chain with nearest-neighbor interactions
\be
H=-t\sum_i(c_{i+1}^{\dag}c_i+c_i^{\dag}c_{i+1})+V\sum_i n_i n_{i+1}.
\ee
This Hamiltonian also preserves the total particle number. Although there might exist strong interaction between the fermions, the above analysis yields the same high-temperature expansion of $\Delta S_2$. Specifically notice that $[\hat{N}_A, n_in_{i+1}]=0$ for all $i$.

Essentially the same analysis leads to the high temperature expansion of $\Delta S_m$, defined as
\begin{multline}
\Delta S_m=S(\rho_m)-S(\rho)=\\
-\lim_{n\to 1} \partial_n {\Tr{\rho_m^n}}-(- \lim_{n\to 1} \partial_n {\Tr{\rho^n}}).
\end{multline}
The high temperature expansion turns out to be
\be
\Delta S_m=\frac{1}{2}(\beta t)^2.
\ee
This result is also immune to the interaction. A similar result has been derived for the mutual information~\cite{nir2020machine}.

\section{CFT methods}
\label{Appendic:cft}
\begin{figure}[h]
	\includegraphics[width=\linewidth]{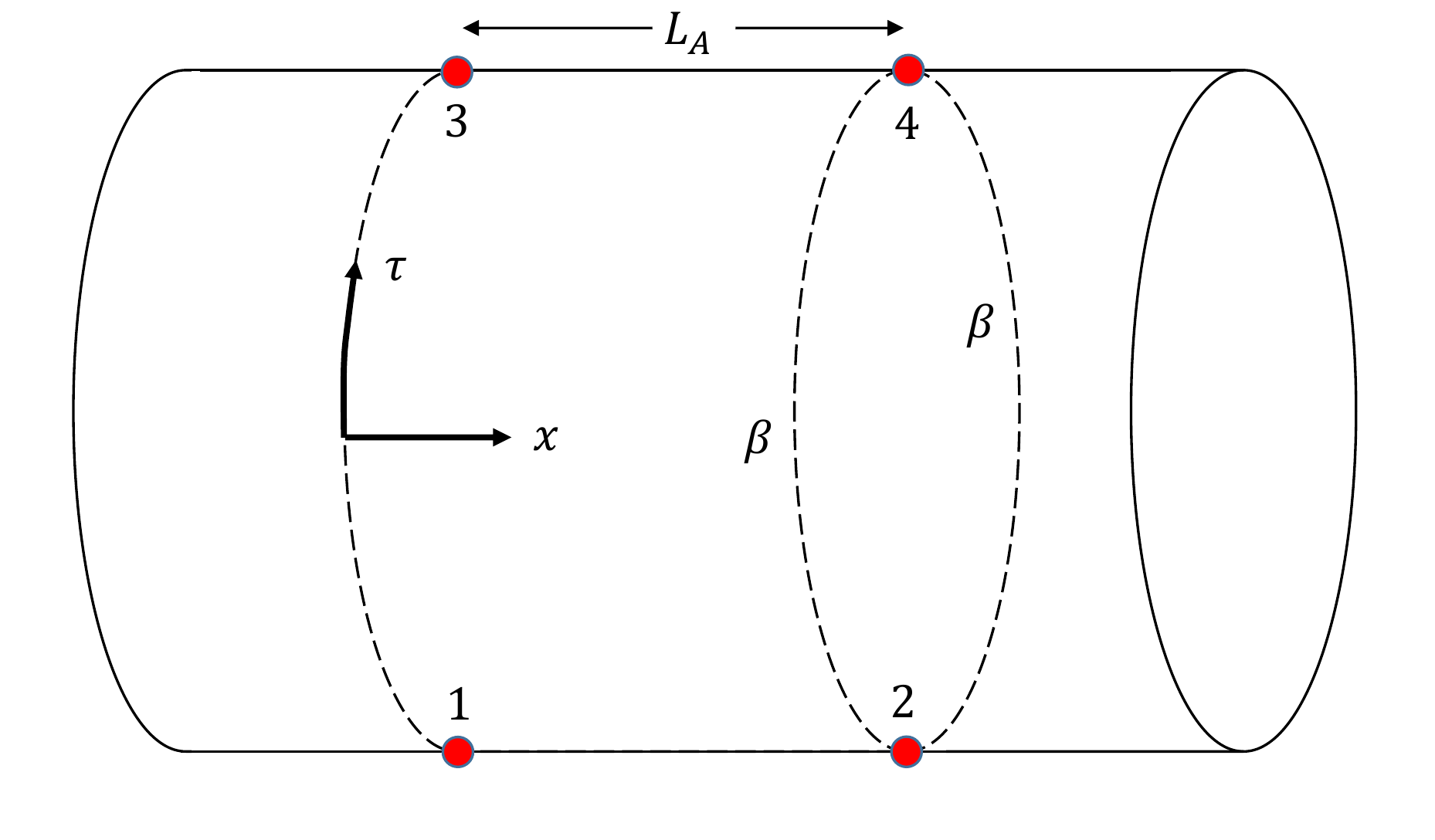}
	\caption{Cylinder geometry of circumference $2 \beta$ where the correlation function $\braket{V_1V_2V_3V_4}$ is computed.
	}
	\label{fig:2cyl}
\end{figure}

In the following, we identify the trace in Eq.~(\ref{eq:trace2beta}) as a path integral on a cylinder of circumference $2\beta$ with angular variable $\tau$ and infinite coordinate $x$, see Fig.~\ref{fig:2cyl}. Following a related computation of the SRE~\cite{goldstein2018symmetry}, each of the operators $e^{\pm i \alpha \hat{N}_A}$ at $\tau=0$ and $\tau=\beta$, is realized by a pair of vertex operator insertions as $e^{i \alpha \hat{N}_A} = e^{i \frac{\alpha}{2 \pi} \phi(\tau,L_A)} e^{-i \frac{\alpha}{2 \pi} \phi(\tau,0)}$. We obtain a normalized correlation function of vertex operators
\be
\label{eq:tracetover}
\frac{\Tr(e^{-\beta H}e^{-i(\alpha_1-\alpha_2) \hat{N}_A}e^{-\beta H}e^{i(\alpha_1-\alpha_2) \hat{N}_A})}{\Tr{e^{-2\beta H}}}=\braket{V_1V_2V_3V_4}.
\ee
Here the 4 vertex operators are located at
\begin{align*}
(0,0)&: ~~V_1 = e^{-\frac{\alpha_1-\alpha_2}{2\pi}\phi(0,0)},\\
(L_A,0)&:~~V_2= e^{\frac{\alpha_1-\alpha_2}{2\pi}\phi( (L_A,0))},\\
(0,\beta)&: V_3=e^{\frac{\alpha_1-\alpha_2}{2\pi}\phi(0,\beta)},\\
(L_A,\beta)&:~~V_4= e^{-\frac{\alpha_1-\alpha_2}{2\pi}\phi(L_A,\beta)}.
\end{align*}

Notice that while  $\Tr{\rho^2}=\frac{\Tr{e^{-2\beta H}}}{Z^2}$, we have
\be
\Tr{\rho_m^2}=\frac{\Tr{e^{-2\beta H}}\int\frac{d\alpha_1d\alpha_2}{(2\pi)^2}\braket{V_1V_2V_3V_4}}{Z^2}.
\ee
The object of interest is entropy change, which can be written as
$\Delta S_2=-\log{\frac{\Tr{\rho_m^2}}{\Tr{\rho^2}}}$ resulting in Eq.~(\ref{eq:tracetover}).
We conclude that $\Delta S_2$ is determined entirely by the 4-point function,
\be
\Delta S_2=-\log{\int\frac{d\alpha_1d\alpha_2}{(2\pi)^2}\braket{V_1V_2V_3V_4}}.
\ee
The 4-point function can be calculated by mapping the cylinder to the complex plane, on which there is a closed formula for the multi-point correlation function. As shown in Appendix~(\ref{sec:correlation functions}), 
\be
\label{resultlastappendix}
\braket{V_1V_2V_3V_4}=\left(\frac{2\beta \tanh{\frac{\pi L_A}{2\beta}}}{\pi}\right)^{-\frac{(\alpha_1-\alpha_2)^2}{\pi^2}}.
\ee

The integration over $\alpha_1$ and $\alpha_2$ can be done exactly.\par
In the following, we consider two temperature regions separately, $T=0$, and $\frac{1}{L_A}\ll T\ll \Lambda$. 

a. $T=0$. In this limit
\be
\lim_{\beta\to\infty}\left(\frac{2\beta \tanh{\frac{\pi L_A}{2\beta}}}{\pi}\right)=L_A.
\ee 
The integration over $\alpha_1$ and $\alpha_2$ gives the aforementioned double log scaling,
\be
\Delta S_2=\frac{1}{2}\log{\log{L_A}}+const.
\ee
\par
b. $\frac{1}{L_A}\ll T\ll \Lambda\sim t$. In this regime $\tanh{\frac{\pi L_A}{2\beta}} \cong 1$ and the 4 point function depends on $\beta$ solely. The same integration gives
\be
\Delta S_2 \sim \frac{1}{2}\log{\log{\frac{const}{T}}}+const'.
\ee
The CFT results hold true only when $T\ll \Lambda$. In Fig.~\ref{fig:comb} of the main text we plot the result of exact integration of Eq.~(\ref{resultlastappendix}) over the $\alpha_i$'s.

\subsection{Derivation of Eq.~(\ref{resultlastappendix})}
\label{sec:correlation functions}
Here we provide the details of the calculation of the 4-point correlation function of vertex operators on cylinder, Eq.~(\ref{resultlastappendix}). 

We introduce complex coordinate $w=x+i\tau$, and map the cylinder to the complex plane by
\be
z=e^{\frac{\pi}{\beta}w}.
\ee
The operators are now inserted at $(1,0), (-1,0), (x, 0), (-x, 0)$, where $(x,y)$ denotes a point on complex plane with $z=x+iy$. Here $x=e^{\frac{\pi}{\beta}L_A}$. 
The multi-point correlation function of vertex operators $V^{A} = e^{iA \phi}$, where $A =\pm \frac{\alpha_i}{2 \pi}$ is given by
\be
\braket{V^{A_1}(z_1,\bar{z}_1)\cdots V^{A_n}(z_n,\bar{z}_n)}=\prod_{i<j}|z_i-z_j|^{2 A_i A_j}.
\ee
According to the above formula, we get
\begin{multline}
\braket{V_1(z_1,\bar{z}_1)V_2(z_2,\bar{z}_2)V_3(z_3,\bar{z}_3)V_4(z_4,\bar{z}_4)}=\\
(x-1)^{-\frac{(\alpha_1-\alpha_2)^2}{\pi^2}}(x+1)^{\frac{(\alpha_1-\alpha_2)^2}{\pi^2}}(2x)^{-\frac{(\alpha_1-\alpha_2)^2}{2\pi^2}}2^{-\frac{(\alpha_1-\alpha_2)^2}{2\pi^2}}.
\end{multline}
To get the correlation function on the cylinder, we have to add the factor associated with the conformal transformation 
\be
\phi^{'}(w,\bar{w})=\left|\frac{\partial z}{\partial w}\right|^{2h}\phi(z,\bar{z}),
\ee
$z=e^{\frac{\pi}{\beta}w}$, $\frac{\partial z}{\partial w}=\frac{\pi}{\beta}z$. We conclude that the 4-point function on the cylinder is given by
\begin{align}
\label{eq:vercorr}
&\quad\braket{V_1(w_1,\bar{w}_1)V_2(w_2,\bar{w}_2)V_3(w_3,\bar{w}_3)V_4(w_4,\bar{w}_4)}\nonumber\\
&=(x-1)^{-\frac{(\alpha_1-\alpha_2)^2}{\pi^2}}(x+1)^{\frac{(\alpha_1-\alpha_2)^2}{\pi^2}}
(2x)^{-\frac{(\alpha_1-\alpha_2)^2}{2\pi^2}}\nonumber\\
&\qquad 2^{-\frac{(\alpha_1-\alpha_2)^2}{2\pi^2}}(\frac{\pi}{\beta}x)^{\frac{(\alpha_1-\alpha_2)^2}{2\pi^2}}(\frac{\pi}{\beta})^{\frac{(\alpha_1-\alpha_2)^2}{2\pi^2}}\nonumber\\
&=(\frac{2\beta}{\pi}\frac{x-1}{x+1})^{-\frac{(\alpha_1-\alpha_2)^2}{\pi^2}}\nonumber\\
&=(\frac{2\beta}{\pi}\tanh{\frac{\pi L_A}{2\beta}})^{-\frac{(\alpha_1-\alpha_2)^2}{\pi^2}}.
\end{align}

\bibliography{refs}
\end{document}